\documentclass[%
 aip,
 jmp,%
 amsmath,amssymb,
 reprint,%
]{revtex4-1}

\usepackage{epsfig}
\usepackage{natbib}
\usepackage{color}
\usepackage{amsmath}
\usepackage{amssymb}

\usepackage{graphicx}
\usepackage{dcolumn}
\usepackage{bm}

\newcommand{\mon}{\begin{displaymath}}
\newcommand{\moff}{\end{displaymath}}

\newcommand{\pd}[2]{\frac{\partial {#1}}{\partial {#2}}}

\newcommand{\eon}{\begin{equation}}
\newcommand{\eoff}{\end{equation}}
\newcommand{\eaon}{\begin{eqnarray}}
\newcommand{\eaoff}{\end{eqnarray}}

\newcommand{\appropto}{\mathrel{\vcenter{
  \offinterlineskip\halign{\hfil$##$\cr
    \propto\cr\noalign{\kern2pt}\sim\cr\noalign{\kern-2pt}}}}}

\newcommand{\kp}{k_\perp}

\newcommand{\kt}{k_\theta}
\newcommand{\ep}{\epsilon_\perp}
\newcommand{\eps}{\epsilon}

\newcommand{\nda}{\dot{n}_\alpha}
\newcommand{\lp}{\left(}
\newcommand{\rp}{\right)}
\renewcommand{\bf}{\mathbf}

\newcommand{\aps}{$\alpha$ particles}
\newcommand{\ap}{$\alpha$-particle}
\renewcommand{\aps}{$\alpha$ particles}
\renewcommand{\ap}{$\alpha$-particle}

\begin{document}

\preprint{AIP/123-QED}

\title[Alpha Channeling with High-field  Launch of Lower Hybrid Waves]{Alpha Channeling with High-field Launch of Lower Hybrid Waves}

\author{I. E. Ochs}
\affiliation{Department of Astrophysical Sciences, Princeton University, Princeton, New Jersey 08540}

\author{N. Bertelli}
\affiliation{Princeton Plasma Physics Laboratory, Princeton, New Jersey 08543}

\author{N. J. Fisch}
\affiliation{Department of Astrophysical Sciences, Princeton University, Princeton, New Jersey 08540}
\affiliation{Princeton Plasma Physics Laboratory, Princeton, New Jersey 08543}

\date{\today}

\begin{abstract}
Although lower hybrid waves are effective at driving currents in present-day tokamaks, they are expected to interact strongly with high-energy particles in extrapolating to reactors.
In the presence of a radial alpha particle birth gradient, this interaction can take the form of wave amplification rather than damping.
While it is known that this amplification more easily occurs when launching from the tokamak high-field side,
the extent of this amplification has not been made quantitative.
Here, by  tracing rays launched from the high-field-side of a tokamak,  the required radial gradients to achieve amplification are calculated for a temperature and density regime consistent with a hot-ion-mode fusion reactor.
These simulations, while valid only in the linear regime of wave amplification, nonetheless  illustrate the possibilities for  wave amplification using high-field  launch of the lower hybrid wave. 
\end{abstract}

\maketitle


\section{\label{sec:level1}Introduction}

The lower hybrid (LH) wave has been shown to be effective in driving plasma currents in tokamaks.\cite{fisch1978confining,fisch87}
Yet there remains  a concern that in a fusion reactor, high-energy $\alpha$ particles born in the plasma core could strongly damp the LH wave. \cite{wong,fisch_92a,wang2014influence}
However, it is possible for a favorable population inversion to appear along the diffusion path, resulting in wave amplification rather than wave damping.\cite{fisch1992interaction}
This amplification effect, which relies on coupling diffusion in energy to diffusion in space,  is often referred to as  \emph{alpha channeling}.

For lower hybrid waves, this amplification appears to be best achieved by launching the  wave from the tokamak high-field side.\cite{ochs2015}
Such a launch is often referred to as ``inside launch," since the magnetic field is strongest near the tokamak hole.
Inside launch represents an engineering challenge, since there is less space near the tokamak hole in which to place waveguides. 
However, there are also several recently-discovered advantages to inside launch, even apart from the opportunity for $\alpha$ channeling, including deeper plasma  penetration of the LH wave and a more protected plasma environment for the LH waveguides.\cite{podpaly2012lower,sorbom2014arc}
When it is launched from the high-field side, the opportunity to amplify the LH wave through interactions with \aps\  adds to these advantages. 
However, while  this amplification indeed appears to more easily occur when launching from the tokamak high-field side,
the extent of this amplification has not yet been made quantitative.
%

To quantify the conditions under which ray amplification can occur, it necessary to consider  the ray paths for the lower hybrid waves, including both damping on electrons and growth on \aps\ along the ray trajectory.
Of course, nonlinear effects can affect importantly the propagation and damping of the lower hybrid waves, particularly with respect to penetration of the hot and dense core of the tokamak reactor.
These nonlinear effects include quasilinear flattening of the distribution function which would lessen the electron damping\cite{cardinali2014self} and parametric effects that might allow the wave to access higher density.\cite{cesario2010current}
The employment of the linear model here is thus not meant to optimize penetration to the plasma core; rather, it is meant to clarify the new physical effects in the simplest model that captures these effects.   It should be noted that the amplification effects due to alpha channeling, to the extent that linear damping by electrons is opposed, will in fact enable deeper penetration of the wave.
In a given tokamak magnetic geometry, with given  density and temperature profiles, the ray paths may be calculated  using  GENRAY.\cite{smirnov2001genray}
For simplicity, near-circular, near-concentric flux surfaces are assumed.
Once the ray trajectories are known, both the damping and the growth can be separately calculated.
%

In the linear approximation for the waves, which for simplicity we adopt here, the effect of the LH waves  on the distribution functions of electrons, ions, or \aps\ is ignored.
Generally in the regimes of interest here, damping on fuel ions is neglected.
In terms of the electron interaction, our linear approximation means that we ignore the formation of a quasilinear plateau, which could lessen the electron damping.
We similarly ignore the relaxation of the \ap\ distribution, which would limit the energy extractable from the \aps.
This linear approximation is thus valid in the limit of low power in the LH wave, where diffusion by waves is small compared to diffusion by collisions, so that the ions and electrons assume Maxwellian velocity distributions, while the high-energy \aps\ assume a slowing-down distribution.

Thus, although limited to low LH wave power, the linear model is used
to demonstrate how inside launch can lead to dramatic $\alpha$-mediated LH wave amplification in the reactor regime at reasonable radial $\alpha$-particle birth gradients.
The very important generalization of these effects to higher powers, or for that matter to more realistic tokamak magnetic configurations,  is outside the scope of this work.
Instead, we demonstrate that the linear damping rate on electrons may be exceeded by the linear growth  rate on \aps, so that the wave undergoes significant convective amplification before it finally is extinguished by linear damping.  
Apart from its own intrinsic interest, the calculation of amplification effects in the low-power limit gives important insights to conditions for amplification in the regime of high-power LH waves, which is the regime where the full potential of the amplification effect would be realized.  

The paper is organized as follows: In Sec.~\ref{sec:linear}, the linear model is explained in greater detail. 
In Sec.~\ref{sec:example}, an example of joint optimization of current drive and $\alpha$ channeling is given.
In Sec.~\ref{sec:discussion}, conclusions and caveats with respect to regimes of applicability of the results are discussed.


\section{\label{sec:linear}Linear Model}

Suppose, for simplicity, a  tokamak, with circular, concentric flux surfaces.
Such a magnetic configuration is conveniently parameterized by the toroidal angle $\phi$, the poloidal angle $\theta$, and the minor radius $r$;
then the flux surface normal vector coincides with the minor radius vector $\hat{r}$.
Since $\alpha$ particle density and temperature tend to be constant over each flux surface, $\alpha$ particle gradients will thus also point along the minor radius vector.

Alpha channeling occurs when a wave with frequency $\omega$ and wavevector $\bf{k}$ encounters an $\alpha$ particle orbiting with velocity $v_\perp$ around a magnetic field $\bf{B}$. 
This interaction couples particle diffusion in energy and space, such that particles that gain energy are pushed in the direction of $\bf{x} = \bf{k} \times \bf{B}$, while those that lose energy are pushed towards $-\bf{x}$.
In a reactor, it is desirable to have $\bf{x}$ point inward along the flux surface normal vector, so that particles lose energy as they are diffused towards the plasma periphery.
Consider now the scalar quantity $\xi$, defined consistently with previous work\cite{ochs2015}, which encodes information about the direction of channeling:
\begin{equation}
\label{xi}
\xi \equiv \frac{\bf{k} \times \bf{B}}{|\bf{k} \times \bf{B}|} \cdot \bf{\hat{r}} \approx - \frac{\kt}{|\kp |},
\end{equation}
where $\kt$ is the $\theta$ component of the wavenumber, and $k_\perp$ is the component perpendicular to the magnetic field.
When $\xi$ is negative, $\alpha$ particles that lose energy will be channeled outwards towards the plasma periphery, as desired.
However, if $|\xi| \ll 1$, the particles are channeled largely poloidally (in the direction of $\hat{\theta}$ rather than radially, so that there is unlikely to be a steep $\alpha$ particle gradient along the direction of channeling.
Thus ideally, $\xi \approx -1$.

This coupling of energetic and spatial diffusion results in diffusion paths in $r-\ep$ space, where $\ep$ is the $\alpha$ particle kinetic energy perpendicular to the magnetic field.	\cite{fisch1992interaction}
For simplicity, we normalize $\eps_\perp$ to the $\alpha$-particle birth energy $\eps_\alpha\approx 3.5$ MeV, so that $\ep=\lp v_\perp/v_\alpha\rp^2$, where 
$v_\alpha$ is the $\alpha$ particle birth energy.
These diffusion paths obey: 
\begin{equation}
\label{diff}
\pd{\ep}{x} = \frac{m_\alpha \Omega_\alpha \omega}{k_\perp \eps_\alpha} = \frac{2 Z_\alpha e B \omega}{k_\perp m_\alpha v_\alpha^2},
\end{equation}
where $Z_\alpha = 2$ is the ion number, $e$ is the electron charge, and $m_\alpha$ is the $\alpha$ particle mass.
%
Since the density may be assumed to be uniformly distributed along the flux surface, only distances along the radial coordinate matter, so that,  using Eq.~(\ref{xi}), we find
\begin{equation}
\pd{\ep}{r} = \frac{1}{\xi} \pd{\ep}{x} = \frac{m_\alpha \Omega_\alpha \omega}{\xi k_\perp \eps_\alpha} = \frac{2 Z_\alpha e B \omega}{\xi k_\perp m_\alpha v_\alpha^2}. \label{eq:dedr}
\end{equation}
Thus, as $\xi$ gets smaller, i.e. as the \aps\ are pushed by the wave more  in the direction of $\theta$, it takes a very large change in energy to move \aps\ even slightly radially.

To calculate the change in power of the wave due to the energetic diffusion of the $\alpha$ particles,
consider an infinitesimal length $l$ of the propagating ray, which we take to have cross-sectional area $A$.
Particles will be pushed in the direction $\bf{x}$, which, since it must be normal to $\bf{k}$, necessarily lies in the cross-sectional plane; we denote the remaining orthogonal direction $\bf{y}$.
Thus, the total number of $\alpha$ particles transferred across a wave section of length $dl$ in a unit time $dt$ is $N_\alpha=J (y dl) dt$, where $J$ is the energetic flux, given from the energetic diffusion coefficient $D_{\eps}$ and the $\alpha$ particle distribution $f(\eps_\perp,x)$ by
\begin{equation}
J = D_\eps \left[ \frac{d}{d\eps} + \lp \frac{dr}{d\eps_\perp} \rp \frac{d}{dx} \right] f(\eps_\perp,x).
\end{equation}
As these particles diffuse, they gain or lose (unnormalized) energy $\Delta \eps_\perp=\eps_\alpha(d \eps_\perp/dx)x$.
Multiplying the number of particles transferred by the energy lost per particle, and integrating across the energetic distribution, we then have (noting $A=xy$): 
\begin{equation}
\frac{dP}{dl} = \int_{0}^{\infty}  d\eps_\perp   \, \eps_\alpha A \lp \frac{d\eps_\perp}{dx} \rp D_\eps \lp \frac{d}{d\eps} + \lp \frac{dx}{d\eps} \rp \frac{d}{dx} \rp f(\eps_\perp,x). \label{eq:p_early}
\end{equation}
To make use of this equation, we must find expressions for the distribution function $f$, diffusion coefficient $D_\eps$, and area $A$.

To calculate the distribution function,  note that the energetic distribution of high-energy $\alpha$ particles on each flux surface is dominated by collisional slowing down on electrons.
Suppose a flux-surface-dependent birth concentration from fusion $\nda(r)$; suppose furthermore that this birth  distribution slows down on electrons, in such a way that that the \aps\ remain on their birth flux surface.  Then the slowing down  distribution is in perpendicular energy velocity space can be put as   
\begin{equation}
f(\ep,r) = \frac{\nda(r)}{2 \nu} \frac{\sqrt{1-\ep}}{\ep} H(1-\eps_\perp), \label{eq:f} 
\end{equation}
where $\nu = 16\sqrt{2\pi m_e} e^4 n_e \times \ln \Lambda/3T_e^{3/2}m_\alpha$ is the collisional damping on electrons and $H(x)$ is the Heaviside function.
As discussed above, throughout this paper,  the LH wave intensities are assumed insufficient to modify the $\alpha$-particle distribution, so that Eq.~(\ref{eq:f}) holds: we will later discuss the consequences of this assumption.

The diffusion coefficient in perpendicular velocity space can be written as \cite{karney,fisch1992interaction} 
\begin{equation}
D_\eps(\ep) = \frac{2 \omega}{\eps_w^3 (\ep - \eps_w)^{1/2}} \lp \frac{v_\text{osc}}{v_\alpha} \rp^2 H(\ep-\eps_w), \label{eq:d}
\end{equation} 
where $\eps_w = (\omega/\kp)^2/v_\alpha^2$, $v_\text{osc} = 2eE/m_\alpha\omega$.
Note that if the perpendicular phase velocity $v_{p,\perp}=\omega/k_\perp$ of the wave is greater than the birth velocity $v_\alpha$ of the $\alpha$ particles, then $\eps_w>1$.
Since $\eps_\perp \leq1$, the diffusion coefficient will then be strictly zero, and there will be no $\alpha$ particle interaction. 

Finally, the power in the wave at a point along the ray trajectory is given approximately by 
\begin{equation}
P = \frac{1}{2} v_g \eps_0 E^2 A, \label{eq:p}
\end{equation} 
where $A$ is the ray's cross-sectional area, $E$ is the electric field, $\eps_0$ is the permittivity constant, and $v_g$ is the group velocity.

Plugging the expressions for the diffusion coefficient (Eq.~\ref{eq:d}), distribution function (Eq.~\ref{eq:f}), and ray power (Eq.~\ref{eq:p}) into Eq.~(\ref{eq:p_early}) allows us to calculate the change in power along the ray due to the $\alpha$ particle interaction:
\begin{align}
\frac{dP}{dl} &= P \lp \frac{Z^2_\alpha e^2 \dot{n}_\alpha}{\nu \eps_0 m \omega \eps_w^3 v_g} \rp \times  \biggl[ - \pi \lp \frac{1-\eps_w}{2 \eps_w^{3/2}} + \frac{1}{2 \sqrt{\eps_w}} \rp \notag\\ & \qquad \qquad + \pi \lp 1 + \eps_w^{-1/2} \rp  \lp 2 \pd{r}{\ep}  \rp  \lp \frac{1}{f}\pd{ f}{r} \rp \biggr]. \label{eq:zerod} 
\end{align}
The two terms within the square brackets have intuitive physical interpretations.
The first term, which is strictly negative, arises from the monotonically decreasing energetic distribution of $\alpha$ particles on each flux surface, which can only result in wave damping.
The second term, arising from the radial $\alpha$ particle birth gradient, can take either sign.
In particular, if the second term is positive and larger in magnitude than the first term, the wave will exponentially amplify on the $\alpha$ particles.

Let us define the radial $\alpha$-particle birth decay length $\lambda_0 \equiv   \nda/( d\nda/dr)$ at which this transition to amplification occurs.
Substituting Eq.~(\ref{eq:dedr}) into Eq.~(\ref{eq:zerod}), and noting $\partial f/ \partial r \approx d \nda/ d r $, yields 
%
\begin{align}
	\lambda_0 =  4 \lp 1-\sqrt{\eps_w} \rp \left(\frac{\omega}{\kp}\right) \left( \frac{\xi}{2 \Omega_\alpha}\right) \propto (1-\sqrt{\eps_w} ) \sqrt{\eps_w}.
	\label{eq:Rfrac}
\end{align}
%
When $\lambda_0$ is negative and the radial decay length $\lambda_\alpha$ in the $\alpha$-particle birth distribution satisfies $\lambda_\alpha < \lambda_0$, then the $\alpha$ particles will amplify, rather than damp, the wave. 

\begin{figure}[b] 
	\center{\includegraphics[width=.95\linewidth]{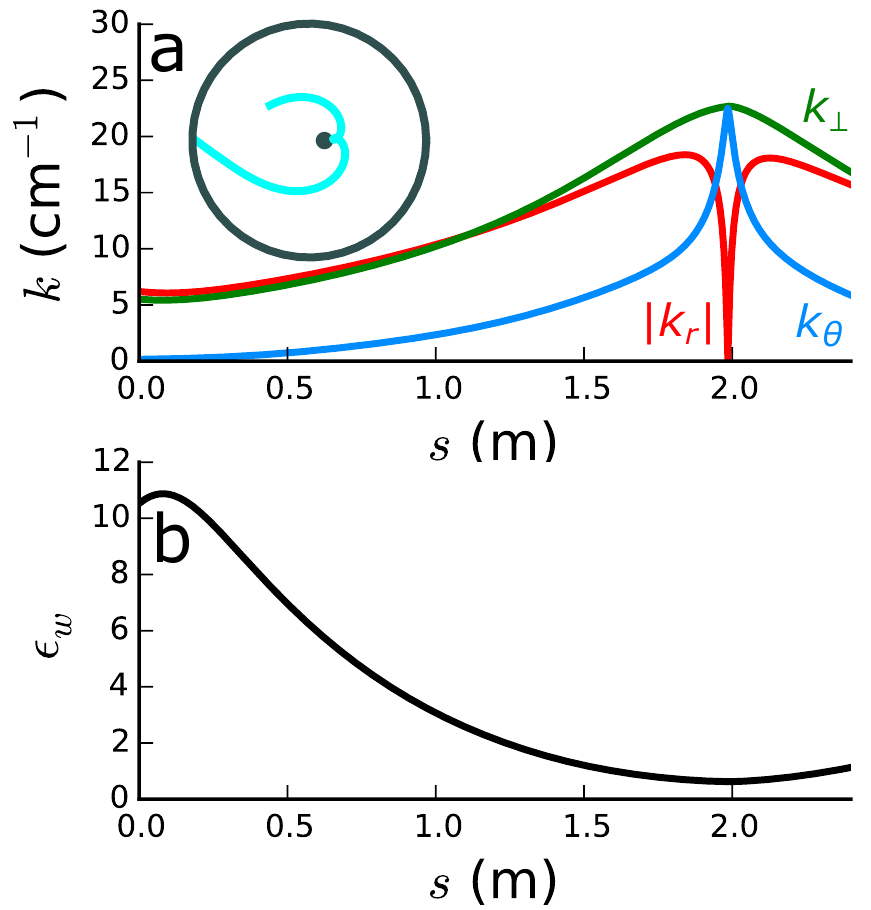}}
	\caption{Simulation results for an ARC-like equilibrium for $n_\parallel=-1.65$, $f=3.7$ GHz, and $\theta = \pi$ (inside launch).
	(a) Inset:  poloidal trajectory of ray. Graph: the evolution of $\kt$ (blue), $| k_r |$ (red), and $k_\perp$ (green)  along the poloidal trajectory length $s$. Near the reflection point ($s \approx 2.1$ m), most of $k_\perp$ is in $k_\theta$, and so $\xi$ becomes large and negative.
	(b) Evolution of $\eps_w \equiv v_{p,\perp}^2/v_\alpha^2$. As the ray gets near the center, the increase in $k_\perp$ (and thus decrease in $v_\perp$) causes the wave to interact with the hot $\alpha$ particles.
	Color available online.}
	\label{fig:k}
\end{figure}

\section{\label{sec:example}Example of Joint Optimization}
As an example of  joint optimization of  the current-drive and $\alpha$-channeling effects in the reactor regime,
consider a nearly circular equilibrium,\cite{bonolicomm}  with parameters similar to the proposed ARC tokamak reactor. \cite{sorbom2014arc}
Accordingly,  take major radius $R = 3.45$m and  minor radius  $a = 1.13$m,
with a toroidal magnetic field of 9.25 T and 7.8 MA of plasma current.
For this equilibrium, the safety factor $q$ monotonically increases with distance from the magnetic axis, from $q$(axis) = 1.015 to $q$(edge) = 2.693.
The  temperature and density profiles for electrons and ions are given analytically on each flux surface by
\begin{align}
	T(r) = \left(T_\text{max} - T_\text{min}\right)\left(1-r^2/a^2\right)^2 + T_\text{min}\\
	n(r) = \left(n_\text{max} - n_\text{min}\right)\left(1-r^2/a^2\right)^2 + n_\text{min},
\end{align}
where  $T_\text{max} =  10$ keV, $T_\text{min} = 500$ eV, $n_\text{max} = 5\times10^{13} \; \text{cm}^{-3}$, and $n_\text{min} = 10^{13} \; \text{cm}^{-3}$.
Here the radial coordinate $r$ refers to the radius of each (essentially) circular flux surface with respect to its own center, \emph{not} with respect to the magnetic axis.
Thus, the density and temperature are constant on a flux surface, where the larger radii surfaces have the smallest densities and temperatures.
The flux surfaces  for this equilibrium exhibit just a small Shafranov shift ($\Delta R < 15$ cm), and,  while not concentric,  are essentially circular.  
Deuterium and tritium ions are in equal proportion. 
The pressure of $\alpha$ particles is neglected.

\begin{figure}[h] 
	\center{\includegraphics[width=.95\linewidth]{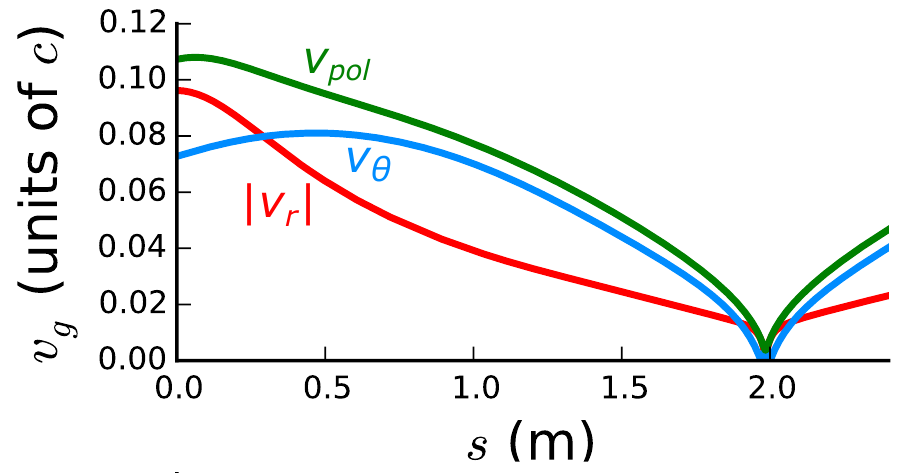}}
	\caption{
	The evolution of the components of the group velocity for an ARC-like equilibrium for $n_\parallel=-1.65$, $f=3.7$ GHz, and $\theta = \pi$ (inside launch). Components are $v_{g,\theta}$ (blue), $| v_{g,r} |$ (red), and $v_{g,\text{pol}} = \sqrt{v_{g,\theta}^2 + v_{g,r}^2}$ (green). Near the reflection point, the poloidal group velocity is dramatically reduced, so that the trajectory spends far ``longer'' where channeling is most favorable.
	Color available online.}
	\label{fig:e_w}
\end{figure}

\begin{figure}[b] 
	\center{\includegraphics[width=.95\linewidth]{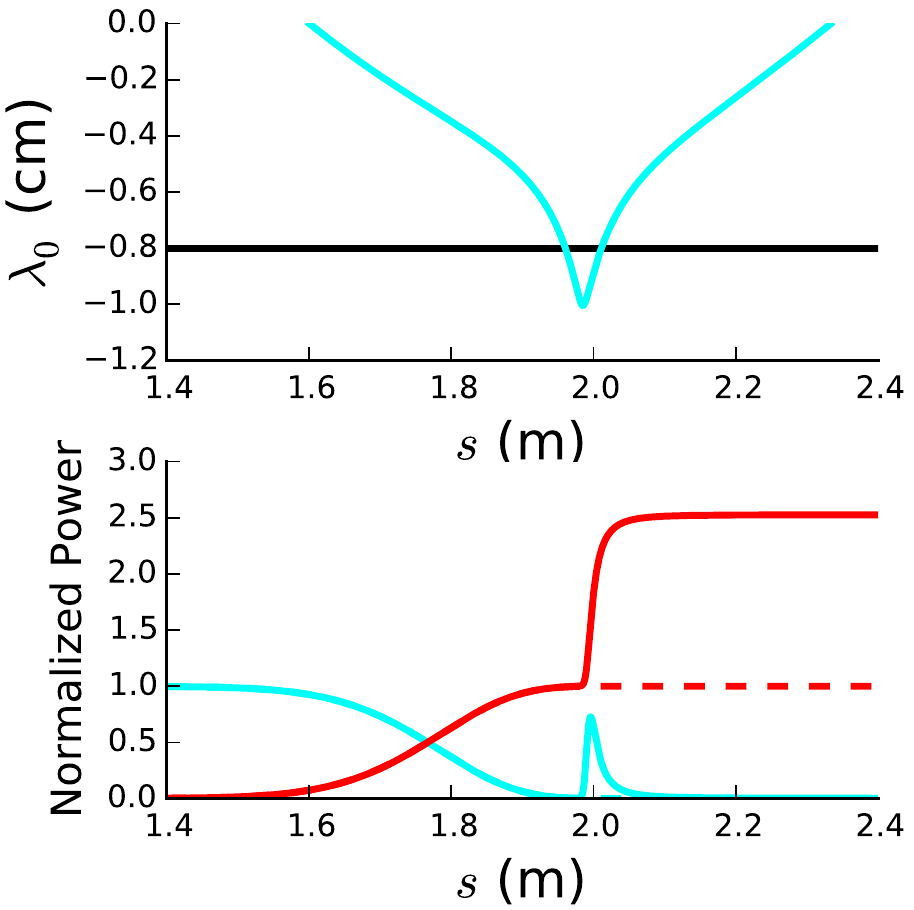}}
	\caption{
	Simulation results near the plasma center for an ARC-like equilibrium for $n_\parallel=-1.65$, $f=3.7$ GHz, and $\theta = \pi$ (inside launch).
	(a) Cyan line: negative radial decay length in cm ($\lambda_0$) of $\alpha$ particles required for zero damping vs. poloidal length along ray. As the ray nears the plasma center ($s\approx 2.0$ m), the maximum allowable decay length increases. Black line:  imposed radial decay length of $\lambda_\alpha=-0.8$ cm. 
	(b) Power remaining in ray (blue) and cumulative power delivered to electrons (red) along ray trajectory, both with (solid) and without (dashed) $\alpha$ particles present. 
	Color available online.}
	\label{fig:supp}
\end{figure}

\begin{figure}[t] 
	\center{\includegraphics[width=1\linewidth]{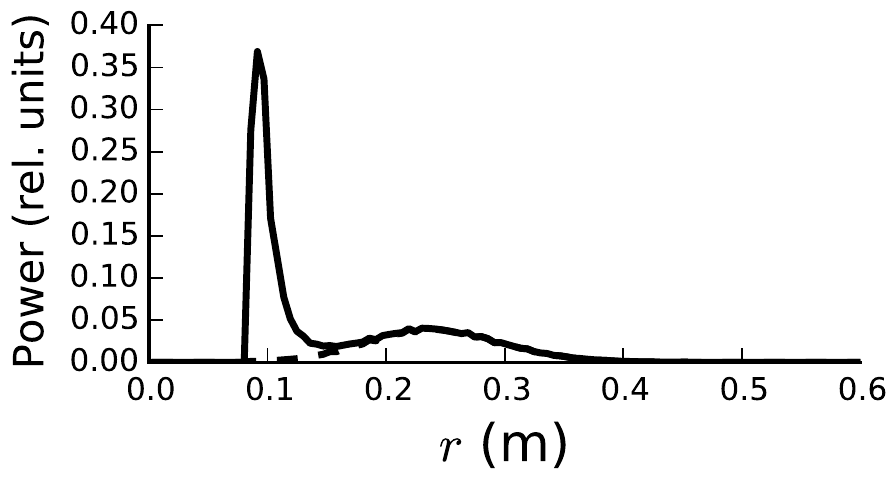}}
	\caption{Radial damping profile on electrons for an ARC-like equilibrium with $n_\parallel=-1.65$, $f=3.7$ GHz, and $\theta = \pi$. 
	Dashed line shows the damping profile in the absence of $\alpha$ particles, solid in the presence of $\alpha$ particles. 
	Most of the power resulting from $\alpha$ channeling is damped near the plasma center.}
	\label{fig:rad}
\end{figure}


With this equilibrium and these kinetic profiles, we calculate one-pass ray trajectories in GENRAY,\cite{smirnov2001genray} a geometrical optics code that also calculates the linear Landau damping on electrons (Landau damping on the ion distribution is assumed to be negligible).\cite{bonoli1984linear,bonoli1986simulation}
Because geometrical optics assumes the WKB (short-wavelength) approximation, the effects of beam width\cite{bertelli2012paraxial} and short-scale inhomogeneity on wave propagation are not captured.
Additionally, the code employs a cold-plasma dispersion relation, so potential warm-plasma effects\cite{wright2014effects} on the ray trajectory are not considered in the current study.
However, both these effects could be incorporated by combining our linear model with full-wave codes\cite{wright2009assessment,meneghini2009full} employing a warm-plasma dispersion relation.

We assume that the ray encounters a constant  $\alpha$-particle gradient,  arising from
an $\alpha$ particle birth distribution with  
decay length $\lambda_\alpha$ 
in the region of wave propagation, and say constant otherwise,   {\it i.e.}
\begin{equation}
	\dot{n}_\alpha (r) = \begin{cases} 
		\dot{n}_{\alpha 0} \exp(-r/\lambda_\alpha) & \text{if } r>r_\text{closest} \\ 
		\dot{n}_{\alpha 0}& \text{otherwise} 
	\end{cases},
\end{equation}
where $r_\text{closest}$ is the distance of closest approach of the ray to the plasma center.
The linear damping (or growth) rate due to  $\alpha$ particles can then be  calculated along the ray trajectory, taking into account the linked diffusion in $r-\ep$ space, according to Eq.~(\ref{eq:dedr}).
Here we see the main simplification of the circular, concentric equilibrium, since the local radial coordinate, with Shafranov-shifted origin, can then be used as the flux normal vector for each surface.

\begin{figure}[b] 
	\center{\includegraphics[width=1\linewidth]{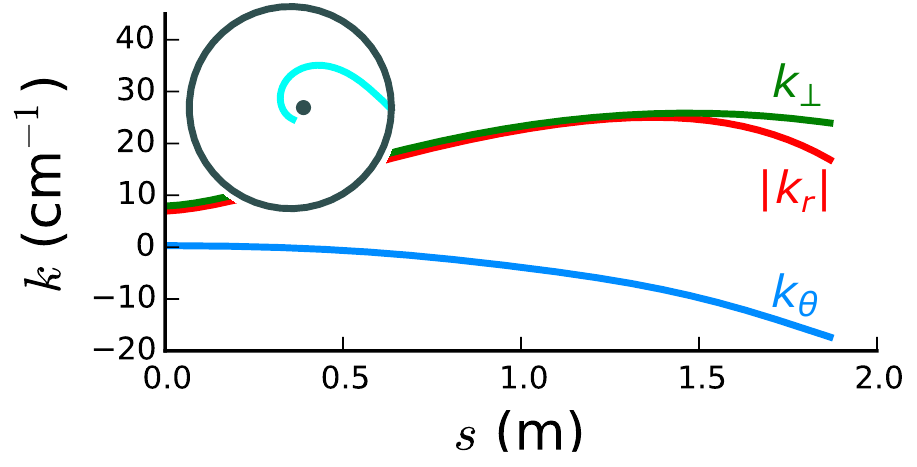}}
	\caption{The evolution of $\kt$ (blue), $| k_r |$ (red), and $k_\perp$ (green)  along the poloidal trajectory length $s$ for low-field-side launch in an ARC-like equilibrium. Launch parameters are $n_\parallel=-2.15$, $f=4.1$ GHz. $k_\theta$ becomes negative rather than positive.}
	\label{fig:bad_kt}
\end{figure}

Consider, for example, a peak high-energy $\alpha$ particle density (above $\eps_l = 30$ keV) of $n_\alpha=\int_{\eps_l}^{\eps_\alpha} f(\ep) d\ep =7.6\times10^{12}\text{ cm}^{-3}$, a 
radial decay length $|\lambda_\alpha| = 0.8$ cm, and  $r_\text{closest} = $ 8.43 cm.
Then consider launching an LH wave from $\theta = \pi$ (inside launch) with $n_\parallel =-1.65$ and $f=3.7$ GHz.  
As this ray gets close to its reflection point near the plasma center, it experiences a sharp increase in $\kt$ and a sharp decrease in $k_r$ (Fig.~\ref{fig:k}a), leading to an increase in the magnitude of $\xi$, the component of channeling normal to the flux surface.
Because $\lambda_0 \propto \xi$, this effect leads to far more favorable conditions for $\alpha$ channeling near the reflection point. 
Crucially, this transfer of $k$ into $k_\perp$ occurs as $k_\perp$ reaches its peak value, causing $\eps_w$ to drop below the interaction threshold at $\eps_w=1$, as seen in Fig.~\ref{fig:k}b.
Thus the wave  interacts most strongly with the $\alpha$ particles precisely at the point where $\lambda_0$ becomes most favorable for channeling.

Simultaneously, the ray also experiences a dramatic decrease in the poloidal group velocity (Fig. \ref{fig:e_w}), causing the ray to spend far longer near the plasma center than suggested by the distance it travels in the poloidal plane, exactly where channeling is most favorable.

\begin{figure}[b] 
	\center{\includegraphics[width=1\linewidth]{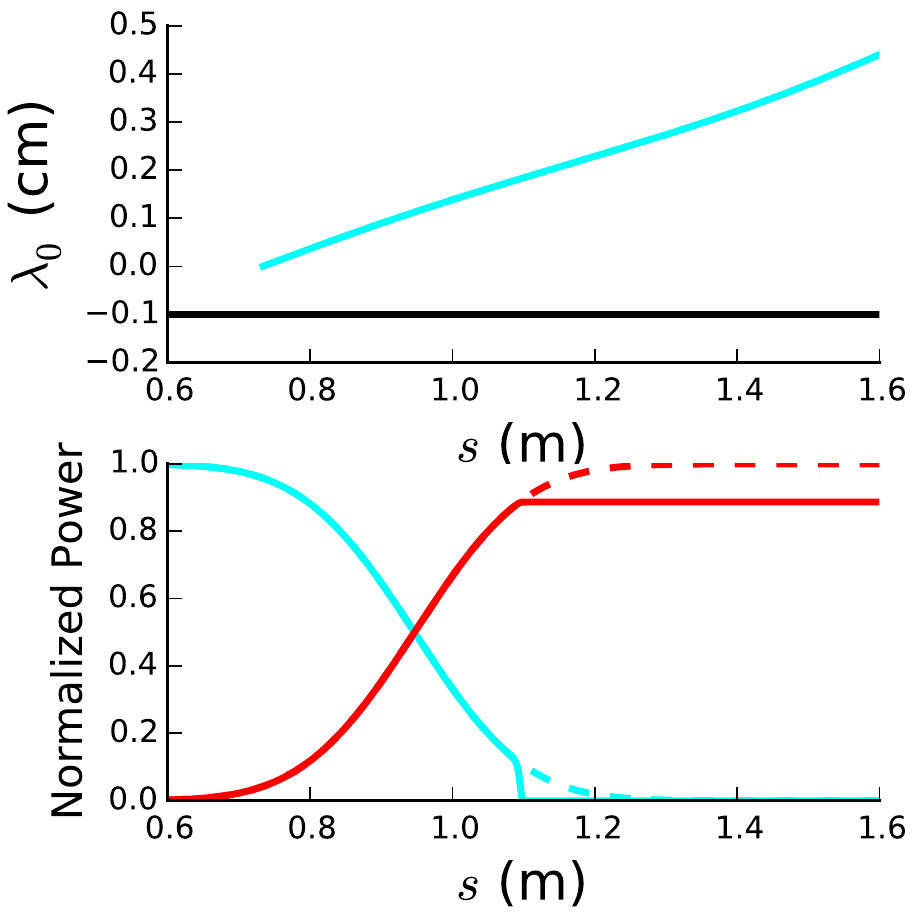}}
	\caption{
	Low-field-side launch simulation results near the plasma center for an ARC-like equilibrium with $n_\parallel=-2.15$, $f=4.1$ GHz. 
	For illustrative purposes only and neglected for the purposes of calculating the ray trajectory, the  $\alpha$ particle distribution has the unrealistically large peak density of $2.18\times10^{141}$ cm$^{-3}$, and a radial decay length of 1mm. (a) $\lambda_\alpha$ (black) and  $ \lambda_0$ (cyan) vs.~$s$, showing $|\lambda_\alpha| < \lambda_0$ for the entire wave-particle interaction region.   (b) Wave power vs.~$s$, showing damping of the wave. 
	Color available online.}
	\label{fig:bad_channeling}
\end{figure}

We can see the importance of this combination of factors in Fig.~\ref{fig:supp}.
Near the plasma center, $\lambda_0$ increases dramatically, resulting in a transition from wave damping to wave amplification as shown in Fig.~\ref{fig:supp}a.
The ray thus regains power from the $\alpha$ channeling effect on the order of its initial power as shown in Fig.~\ref{fig:supp}b.  
As a result, the absorbed power by the electrons is more than twice the launched power in the ray.
Furthermore, this excess power is all damped on electrons near the plasma center (see Fig. \ref{fig:rad}), which is more favorable for plasma stability.

It was previously found from essentially analytic considerations that launching from the high-field side, although unconventional, would be  critical to achieve the wavenumbers needed for $\alpha$ channeling.\cite{ochs2015}
Our present simulations confirm this finding.
As can be seen in Fig.~\ref{fig:bad_kt}. wave launch from the low-field side results in large negative values of $k_\theta$ near the plasma center.
This reversal in the poloidal wavenumber means that, although $\lambda_0$ was large enough in \emph{magnitude} for wave amplification, it had the wrong \emph{sign}, as shown in Fig.~\ref{fig:bad_channeling}a.
With the wrong sign of $k_\theta$, the $\alpha$ particle would actually have to have higher concentration on the plasma periphery than the plasma center for amplification to occur.
Thus, even when the most favorable possible conditions are imposed in terms of the magnitude of the \ap\ gradient, namely  $|\lambda_\alpha| < |\lambda_0|$, the wave is strictly damped rather than amplified on the $\alpha$ particles.
To accentuate this point, we show in Fig.~\ref{fig:bad_channeling}b  how even in the case of an unrealistically and unphysically high peak $\alpha$ particle density, the wave is damped on the \aps\ in the case of high-field launch. 
Thus, it is the persistence of negative $k_\theta$ in launching from the low-field side that necessitates high-field-side launch for LH-mediated $\alpha$ channeling.

\begin{figure}[t] 
	\center{\includegraphics[width=1\linewidth]{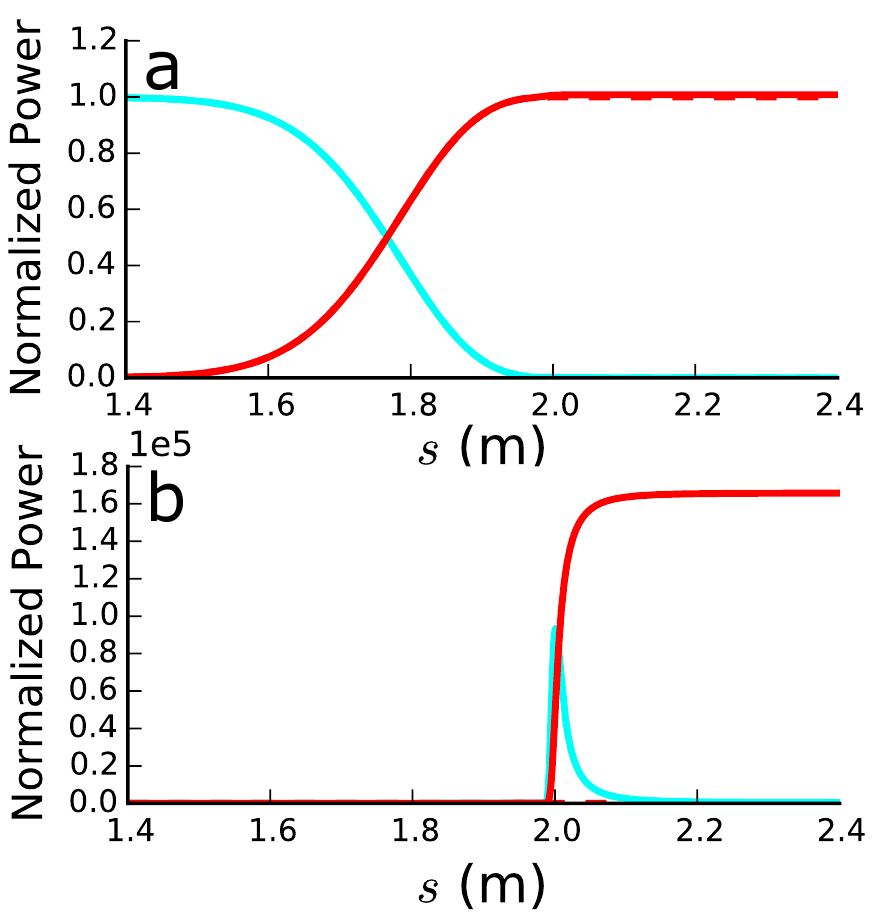}}
	\caption{
	Effect of factor-of-three decrease (a) and increase (b) in $\alpha$ particle concentration on wave amplification for an ARC-like equilibrium for $n_\parallel=-1.65$, $f=3.7$ GHz, and $\theta = \pi$ (inside launch).
	Note change in scale between (a) and (b). 
	Color available online.}
	\label{fig:linear_lims}
\end{figure}

Note also that, in the case of amplification, there is great sensitivity to the number of \aps, since, in the linear limit, amplification is exponential-like, with the amplification rate is directly proportional to the number of \aps\ participating.   
To illustrate this sensitivity, consider, for example,   a factor-of-three decrease  in the $\alpha$ particle density for the same ray as in Fig.~\ref{fig:supp}b.
This results in effectively extinguishing the amplification effect, as shown in Fig.~\ref{fig:linear_lims}a.
On the other hand, a factor-of-three increase in the $\alpha$ particle density for this same ray 
results in a factor of $1.6\times10^5$ increase in wave power, as shown in Fig.~\ref{fig:linear_lims}b.
Of course, for any initially substantive power level, the amplification exhibited in Fig.~\ref{fig:linear_lims}b  is likely to violate the assumption of linearity in the rf power, since at high intensities, both the electron and \ap\ distribution functions would respond to the waves.
However, this example indicates how in principle large effects could take place, which for accuracy would necessitate consideration of nonlinear or quasilinear effects.

\section{\label{sec:discussion}Summary and Discussion}

What has been shown through the simulations presented here is that LH waves launched from the tokamak high-field side, which are already predicted to be advantageous from an engineering standpoint,\cite{podpaly2012lower,sorbom2014arc} can experience dramatic amplification as they penetrate near the plasma center.  
The convectively amplified wave eventually damps on electrons, but in the process it takes energy from the \aps\  leaving it with more energy with which to drive current.
Effectively, the current-drive efficiency is increased, possibly dramatically, if the \ap\ gradients are large enough. 
Thus, the effect could prove critical to achieving efficient current drive at the plasma core.

The effect occurs as follows:  
For high-field side launch, the poloidal wavenumber has the proper sign for $\alpha$ channeling.  
Then, as the LH wave encounters its deepest penetration, it circles the plasma core, meaning that its radial group velocity necessarily vanishes, leaving the full perpendicular wavenumber in the poloidal direction. 
Since for the LH wave, the perpendicular wavenumber is far greater than its parallel wavenumber, the poloidal wavenumber thus grows large enough to capitalize on radial gradients in the \ap\ distribution.  
At the same time as the radial group velocity vanishes, the poloidal group velocity is minimized, so that the ray spends more time just at the fortuitous location where the channeling effect is thus optimized.
The confluence of these circumstances suggests that efficient channeling can occur for inside launch, provided that the ray encounters a sufficiently steep radial distribution of energetic \aps\ at its point of closest approach to the plasma core. 
In practice, the choice of launch parameters might therefore be dictated by the location of the steepest \ap\ gradient. 
In the example given, for a reasonable total number of \aps, and a steep gradient, the wave power damped on electrons could be increased by a factor of 2.53, with most of that damping occurring near the plasma core. 
%

However, there are several caveats worth noting.

First, our model assumed that the RF waves were not strong enough to significantly modify the $\alpha$-particle distribution.
While valid in the limit of weak RF, this model is insufficient in the main cases of interest, namely when the diffusion of \aps\ by the wave  
occurs on a time scale faster than the slowing down of the \aps.  
For if this were not so, then necessarily most of the \ap\ energy would have been lost to collisions before the waves could capture that energy.
To achieve significant savings in a reactor, such as to support in full the plasma current or to achieve the hot ion mode (where the ion temperature exceeds the electron temperature), the lower hybrid waves would be in a regime in which the linear approximation fails.
Similarly, it would not be possible to apply directly these results to the cyclic operation regime of current drive,\cite{fisch_transformer} where synergies were recognized in achieving current drive together with $\alpha$ channeling, again because the cyclic operation regime would be far from the linear regime where these results are valid. 
Note also that the regime of intense LH waves would also necessitate a quasilinear treatment of the electrons as well. 

Second, the wave amplification relies on a favorable population inversion, which requires for lower hybrid waves that the $\alpha$ particle gradients be on the order of centimeters.
For preliminary reactor experiments, sustained fusion conditions are only marginally met in the plasma center, and not at all nearby, which might indeed result  in  a peaked  enough energetic \ap\ distribution.
However, in an advanced reactor, steep gradients are more difficult,  and the \ap\ distribution may be flattened along the diffusion paths when the waves are intense enough that a significant fraction of the  alpha-particle energy is extracted.\cite{fisch_94}  
In that case, the $\alpha$ particles would not slow down significantly through collisions before being diffused to lower energy by the LH waves.  
To maintain the sharp spatial gradients, therefore, a second wave might be used in addition to the LH wave.\cite{fisch_twowave}  
An example of a second wave that might be useful is one that pushes the \aps\ further in space for the same amount of energy extracted, such as the ion Bernstein wave,\cite{valeo1994excitation,fisch_95a} possibly even together with waves of even lower frequency such as toroidal Alfven eigenmodes.\cite{herrmann_97}.  
Note that the use here of a second wave is essentially different from other uses of a second wave to optimize the current drive efficiency,\cite{Fidone84, Fidone04, Dumont04, Giruzzi04, Rosa, farina, Dnestrovskij,chen2012,Huang} since, rather than  promoting absorption by higher velocity electrons,  the second wave  facilitates the absorption of energy to the LH wave from the $\alpha$ particles. Both methods should lead to an effective increase in the efficiency.

In sum, although there remain several caveats in applying these results to the most useful regimes, namely those in which the \ap\ amplification leads to very significant channeling of energy, the linear regime addressed here is strongly indicative of the potential inherent in LH waves launched from the high-field side.


\begin{acknowledgments}
This work was performed under U.S.~DOE contract DE-AC02-09CH11466. The authors are grateful to P.~Bonoli for providing a GENRAY input file for an ARC-like equilibrium.  One of us (IEO) thanks the support of the National Undergraduate Fellowship Program in Plasma Physics and Fusion Energy Sciences.

\end{acknowledgments}


\bibliography{ref5_njf,more_coupling}

\clearpage
\newpage

\end{document}